\documentclass[5p,twocolumn]{elsarticle}
\usepackage{amsmath,amssymb}
\usepackage{graphicx}

\newcommand{\eqdef}{\stackrel{\text{def}}{=}}
\newcommand{\n}{\nonumber\\}
\newcommand{\bm}{\boldsymbol}
\newcommand{\ai}{\text{I}}
\newcommand{\ait}{\text{II}}
\newcommand{\cF}{c_{\text{\tiny$\mathcal{F}$}}}
\newcommand{\ignore}[1]{}
\newcommand{\Romannumeral}[1]{\uppercase\expandafter{\romannumeral#1}}

\allowdisplaybreaks[4]

\journal{Physics Letters B}

\begin{document}

\begin{frontmatter}

%%%%%%%%%%%%%%%%%%%%%%%%%%%%%%%%%%%%%%%%%%%%%%%%%%%%%%%%%%%%
%                                                          %
%  Title page                                              %
%                                                          %
%%%%%%%%%%%%%%%%%%%%%%%%%%%%%%%%%%%%%%%%%%%%%%%%%%%%%%%%%%%%
\title{Exactly Solvable Quantum Mechanics and Infinite Families of
Multi-indexed Orthogonal Polynomials}

\author[SO]{Satoru Odake\corref{cSO}}
\ead{odake@azusa.shinshu-u.ac.jp}
\author[RS]{Ryu Sasaki}
\address[SO]{Department of Physics, Shinshu University,
     Matsumoto 390-8621, Japan}
\address[RS]{Yukawa Institute for Theoretical Physics,
     Kyoto University, Kyoto 606-8502, Japan}
\cortext[cSO]{Corresponding author.}

\begin{abstract}
Infinite families of {\em multi-indexed} orthogonal polynomials are
discovered as the solutions of exactly solvable one-dimensional quantum
mechanical systems. The simplest examples, the one-indexed orthogonal
polynomials, are the infinite families of the {\em exceptional} Laguerre
and Jacobi polynomials of type I and II constructed by the present authors.
The totality of the integer indices of the new polynomials are finite
and they correspond to the degrees of the `{\em virtual state}
wavefunctions' which are `deleted' by the generalisation of Crum-Adler
theorem. Each polynomial has another integer $n$ which counts the nodes.
\end{abstract}


%%%%%%%%%%%%%%%%%%%%%%%%%%%%%%%%%%%%%%%%%%%
% PACS 2006
% 02.       Mathematical methods in physics
%   02.20.Uw  Quantum Groups
%   02.30.Gp  Special functions
%   02.30.Ik  Integrable systems
% 03.       Quantum mechanics, field theories, and special relativity
%   03.65.-w  Quantum mechanics
%   03.65.Ca  Formalism
%   03.65.Fd  Algebraic methods
%   03.65.Ge  Solutions of wave equations: bound states
% 45.       Classical mechanics of discrete systems
%   45.20.-d  Formalisms in classical mechanics
%%%%%%%%%%%%%%%%%%%%%%%%%%%%%%%%%%%%%%%%%%%

%%%%%%%%%%%%%%%%%%%%%%%%%%%%%%%%%%%%%%%%%%%
% PACS 2008
% 02.00.00 Mathematical methods in physics
%   02.30.Gp Special functions
%   02.30.Ik Integrable systems
% 03.00.00 Quantum mechanics, field theories, and special relativity
%   03.65.-w Quantum mechanics
%   03.65.Ca Formalism
%   03.65.Db Functional analytical methods
%   03.65.Fd Algebraic methods
%   03.65.Ge Solutions of wave equations: bound states
% 45.00.00 Classical mechanics of discrete systems
%   45.20.-d Formalisms in classical mechanics
%%%%%%%%%%%%%%%%%%%%%%%%%%%%%%%%%%%%%%%%%%%

\end{frontmatter}
%%%%%%%%%%%%%%%%%%%%%%%%%%%%%%%%%%%%%%%%%%%%%%%%%%%%%%%%%%%%%%%
%                                                             %
%  1. Introduction                                            %
%                                                             %
%%%%%%%%%%%%%%%%%%%%%%%%%%%%%%%%%%%%%%%%%%%%%%%%%%%%%%%%%%%%%%%
\section{Introduction}
\label{intro}
\setcounter{equation}{0}

We will report on the discovery of infinite families of {\em multi-indexed} 
orthogonal polynomials, which form the main parts of the eigenfunctions of
exactly solvable one-dimensional quantum mechanical systems.
This will add a huge number of new members to the inventory of exactly
solvable quantum mechanics \cite{infhul,susyqm}, which had seen rapid
increase due to the recent discovery of various kinds of infinite families
of {\em exceptional} orthogonal polynomials \cite{os16}--\cite{os24}.
The method for deriving the multi-indexed orthogonal polynomials is simple
generalisation of the well-known method of Crum \cite{crum} and its
modification by Krein-Adler \cite{adler}.
Starting from an exactly solvable Hamiltonian system, Crum's method
provides another exactly solvable system by successively deleting the
lowest lying eigenstate. The modification due to Krein-Adler uses deletion
of a finite number of higher eigenstates satisfying certain conditions.
The present method is based on the same spirit of `deleting' a finite
number of {\em virtual states} instead of the eigenstates. Here the
virtual state wavefunctions are the solutions of the Schr\"{o}dinger equation
of the original system and they have  {\em negative energies}, with respect
to the groundstate eigenvalue. The virtual state wavefunctions have
{\em no node in the interior} and they are {\em square-non-integrable}.
Their reciprocals are {\em square-non-integrable}, too.
We apply this method to the Hamiltonian systems of the radial oscillator
potential and the Darboux-P\"oschl-Teller potential, which are the best
known examples of exactly solvable quantum mechanics \cite{infhul,susyqm}.
Their eigenfunctions consist of the Laguerre and the Jacobi polynomials,
respectively.

When only one virtual state of degree $\ell$ ($\ge1$) is deleted,
the resulting system is the exceptional $X_\ell$ Laguerre or Jacobi
polynomials of type I or II \cite{os16,os19}.
In general $M$ distinct virtual states of degree $\text{v}_j$ ($\ge1$)
are deleted and the resulting eigenfunctions consist of orthogonal
polynomials indexed by the integers $(\text{v}_1,\ldots,\text{v}_M)$.
The multi-indexed orthogonal polynomials are natural generalisation of
the exceptional orthogonal polynomials. They start at degree $\ell\ge1$
instead of a degree 0 constant term, thus avoiding the constraints by
Bochner's theorem \cite{bochner,gomez}.
As solutions of exactly solvable quantum mechanical systems they constitute
a complete system of orthogonal functions. The first examples of the
exceptional Laguerre and Jacobi polynomials are constructed by
G\'{o}mez-Ullate et al.\ \cite{gomez} in the framework of Sturm-Liouville
theory and by Quesne \cite{quesne} within the shape invariant \cite{genden}
and exactly solvable quantum mechanics. They are equivalent to the lowest
members ($\ell=1$) of the $X_\ell$ Laguerre and Jacobi polynomials derived
in \cite{os16} by the present authors.
The present method could be considered as the generalisation of the
Darboux-Crum transformations \cite{darb} applied for the derivation of
the exceptional orthogonal polynomials \cite{junkroy}--\cite{os20}.

It is expected that these new polynomials would play important roles
in physics, mathematics and related disciplines. The multi-indexed
Jacobi polynomials provide infinitely many global solutions of second
order Fuchsian differential equations with $3+\ell$ regular singularities
\cite{hos}.
Other theoretical developments are:
non-linear identities underlying the shape invariance \cite{os18} are
clarified, the structure of the extra zeros of the new orthogonal
polynomials \cite{hs4} are exemplified, two-step Darboux transformations
\cite{gomez3} are explicitly constructed for the type II exceptional
Laguerre polynomials, various use of non-physical states \cite{ho2,Grandati}
for generating the exceptional polynomials are suggested and several
applications \cite{midyaroy,ho} are reported. In this connection, let us
mention references of multi-step Darboux-Crum transformations, higher
order susy transformations, $\mathcal{N}$-fold susy transformations, etc
\cite{Nsusy}. Some of them utilise non-physical solutions.

This Letter is organised as follows.
The general scheme of virtual states deletion is presented in section two.
The explicit forms of the new polynomials are derived in section three,
together with the forward-backward shift operators and the second order
differential operators. The final section is for a summary and comments.

%%%%%%%%%%%%%%%%%%%%%%%%%%%%%%%%%%%%%%%%%%%%%%%%%%%%%%%%%%%%%%%
%                                                             %
%  2. Adler-Crum Scheme for Virtual States Deletion           %
%                                                             %
%%%%%%%%%%%%%%%%%%%%%%%%%%%%%%%%%%%%%%%%%%%%%%%%%%%%%%%%%%%%%%%
\section{Adler-Crum Scheme for Virtual States Deletion}
\label{sec:virt}

Here we present the method of virtual states deletion in its generic and
formal form. For rigorous treatments of possible singularities at the
boundaries, we need to specify the details of the Hamiltonian systems,
which will be done in the next section.
Throughout this paper we choose all the solutions of the Schr\"{o}dinger
equations to be real.
We start with an exactly solvable one dimensional quantum mechanical
system defined in an interval $x_1<x<x_2$,
\begin{align}
  \mathcal{H}\phi_n(x)&=\mathcal{E}_n\phi_n(x)\quad
  (n\in\mathbb{Z}_{\ge0}),
  \label{sheq}\\
  (\phi_m,\phi_n)&\eqdef\int_{x_1}^{x_2}\!\!dx\,\phi_m(x)^*\phi_n(x)
  =h_n\delta_{m\,n}.
\label{orth}
\end{align}
For simplicity, we assume that it has discrete eigenvalues only with
vanishing groundstate energy:
$0=\mathcal{E}_0<\mathcal{E}_1<\mathcal{E}_2<\cdots,$
which allows us to express the potential in terms of the groundstate
wavefunction $\phi_0(x)$ which has no node:
\begin{equation}
  \mathcal{H}=-\frac{d^2}{dx^2}+U(x),\quad
  U(x)=\frac{\partial_x^2\phi_0(x)}{\phi_0(x)}.
\end{equation}
Let us suppose that the Schr\"{o}dinger equation has a finite or infinite
number of {\em virtual state} solutions with {\em negative energies}:
\begin{equation} 
  \mathcal{H}\tilde{\phi}_\text{v}(x)
  =\tilde{\mathcal{E}}_\text{v}\tilde{\phi}_\text{v}(x),\quad
  \tilde{\mathcal{E}}_\text{v}<0\quad(\text{v}\in\mathcal{V}),
\end{equation}
which {\em have no node} in the interior $x_1<x<x_2$.
Here $\mathcal{V}$ denotes the index set of the virtual states.
We also require the infinite norm conditions,
$(\tilde{\phi}_\text{v},\tilde{\phi}_\text{v})
=(1/\tilde{\phi}_\text{v},1/\tilde{\phi}_\text{v})=\infty$ and this
means that the virtual state wavefunctions vanish at one boundary and
diverge at the other:
\begin{align}
  \lim_{x\to x_j}\tilde{\phi}_\text{v}(x)
  &=\lim_{x\to x_j}\partial_x\tilde{\phi}_\text{v}(x)=0,\n
  \lim_{x\to x_k}\tilde{\phi}_\text{v}(x)
  &=\lim_{x\to x_k}\partial_x\tilde{\phi}_\text{v}(x)=\infty,
  \label{vbc}
\end{align}
where $j\neq k\in\{1,2\}$ and $\forall \text{v}\in\mathcal{V}$.
The set of eigenfunctions $\{\phi_n(x)\}$ forms a complete set in an
appropriate Hilbert space $\mathsf{H}$, for example:
\begin{align*}
  \mathsf{H}&\eqdef\bigl\{\psi(x)\in L^2[x_1,x_2]\bigm|
  \lim_{x\downarrow x_1}(x-x_1)^{\frac12}\partial_x\psi(x)=0,\n
  &\hspace*{36mm}
  \lim_{x\uparrow x_2}(x_2-x)^{\frac12}\partial_x\psi(x)=0\bigr\}.
\end{align*}
(For $x_1=-\infty$ and $x_2=\infty$, replace
$(x-x_1)^{\frac12}\to(-x)^{\frac12}$ and $(x_2-x)^{\frac12}\to x^{\frac12}$.)
That is any function in $\mathsf{H}$ which is orthogonal to all the
eigenfunctions has a zero norm:
\begin{equation*}
  (\Psi,\phi_n)=0\quad(\forall n\in\mathbb{Z}_{\ge0})
  \ \ \Rightarrow\ \ (\Psi,\Psi)=0.
\end{equation*}

After $M$-steps of deleting the lowest lying eigenstates by Crum's method
\cite{crum}, the resulting Hamiltonian system is iso-spectral to the
original one given by
\begin{align*}
  &\mathcal{H}^{[M]}\phi_n^{[M]}(x)=\mathcal{E}_n\phi_n^{[M]}(x)\quad
  (n=M,M+1,\ldots),\\
  &\phi_n^{[M]}(x)\eqdef
  \frac{\text{W}[\phi_0,\phi_1,\ldots,\phi_{M-1},\phi_n](x)}
  {\text{W}[\phi_0,\phi_1,\ldots,\phi_{M-1}](x)},\\
 & (\phi_m^{[M]},\phi_n^{[M]})
  =\prod_{j=0}^{M-1}(\mathcal{E}_n-{\mathcal E}_{j})\cdot h_n\delta_{m\,n},\\
  &U^{[M]}(x)\eqdef U(x)-2\partial_x^2\log
  \bigl|(\text{W}[\phi_0,\phi_1,\ldots,\phi_{M-1}](x)\bigr|,
  \nonumber
\end{align*}
in which $\text{W}[f_1,\cdots,f_n](x)\eqdef\det\bigl(\partial_x^{j-1}
f_k(x)\bigr)_{1\le j,k\le n}$ is the Wronskian.
These formulas have been repeatedly rediscovered under various names,
higher derivative susy, $\mathcal{N}$-fold susy, etc \cite{Nsusy}.
The corresponding Adler-Crum formulas for deleting the set of $M$ distinct
eigenstates specified by $\mathcal{D}\eqdef\{d_1,d_2,\ldots,d_M\}$,
$d_j\ge0$, are
\begin{align*}
  &\mathcal{H}^{[M]}\phi_n^{[M]}(x)=\mathcal{E}_n\phi_n^{[M]}(x)\quad
  (n\in\mathbb{Z}_{\ge0}\backslash\mathcal{D}),\\
  &\phi_n^{[M]}(x)\eqdef
  \frac{\text{W}[\phi_{d_1},\phi_{d_2},\ldots,\phi_{d_M},\phi_n](x)}
  {\text{W}[\phi_{d_1},\phi_{d_2},\ldots,\phi_{d_M}](x)},\\
 & (\phi_m^{[M]},\phi_n^{[M]})
  =\prod_{j=1}^M(\mathcal{E}_n-{\mathcal E}_{d_j})\cdot h_n\delta_{m\,n},\\
  &U^{[M]}(x)\eqdef U(x)-2\partial_x^2\log
  \bigl|\text{W}[\phi_{d_1},\phi_{d_2},\ldots,\phi_{d_M}](x)\bigr|.
  \nonumber
\end{align*}
The set $\mathcal{D}$ must satisfy the conditions $\prod_{j=1}^M(m-d_j)\ge0$,
$\forall m\in\mathbb{Z}_{\ge 0}$ \cite{adler,gos}.
The exactly solvable Hamiltonian system obtained by deleting $M$ distinct
{\em virtual states} $\mathcal{D}=\{d_1,\ldots,d_M\}$, $d_j\in\mathcal{V}$
looks almost the same:
\begin{align}
  &\mathcal{H}^{[M]}\phi_n^{[M]}(x)=\mathcal{E}_n\phi_n^{[M]}(x)\quad
  (n\in\mathbb{Z}_{\ge0}),
  \label{mnham}  \\
  &\mathcal{H}^{[M]}\tilde{\phi}_\text{v}^{[M]}(x)
  =\tilde{\mathcal{E}}_\text{v}\tilde{\phi}_\text{v}^{[M]}(x)\quad
  (\text{v}\in\mathcal{V}\backslash\mathcal{D}),
  \label{mnvham}\\
  &\phi_n^{[M]}(x)\eqdef
  \frac{\text{W}[\tilde{\phi}_{d_1},\tilde{\phi}_{d_2},\ldots,
  \tilde{\phi}_{d_M},\phi_n](x)}
  {\text{W}[\tilde{\phi}_{d_1},\tilde{\phi}_{d_2},\ldots,
  \tilde{\phi}_{d_M}](x)},\\
  &(\phi_m^{[M]},\phi_n^{[M]})
  =\prod_{j=1}^M(\mathcal{E}_n-\tilde{\mathcal{E}}_{d_j})\cdot
  h_n\delta_{m\,n},
  \label{fineig}\\
  &\tilde{\phi}_\text{v}^{[M]}(x)\eqdef
  \frac{\text{W}[\tilde{\phi}_{d_1},\tilde{\phi}_{d_2},\ldots,
  \tilde{\phi}_{d_M},\tilde{\phi}_\text{v}](x)}
  {\text{W}[\tilde{\phi}_{d_1},\tilde{\phi}_{d_2},\ldots,
  \tilde{\phi}_{d_M}](x)},
  \label{finvirt}\\
  &U^{[M]}(x)\eqdef U(x)-2\partial_x^2\log
  \bigl|\text{W}[\tilde{\phi}_{d_1},\tilde{\phi}_{d_2},\ldots,
  \tilde{\phi}_{d_M}](x)\bigr|.\!\!
  \label{finpot}
\end{align}
It should be stressed that the final results after $M$ deletions are
independent of the orders of deletions.
The algebraic side of the derivation of \eqref{mnham}--\eqref{finpot} is
basically the same as that for the Adler-Crum case.
The essential difference is the verification of the inheritance at
each step of the properties of the virtual states; no-nodeness and
the proper boundary behaviours.
This is guaranteed by choosing the parameters of the theory $g$ and $h$
large enough, see \eqref{Lbound} and \eqref{Jbound}.

Here we show the process $\mathcal{H}\to\mathcal{H}^{[1]}$.
It is elementary to show
\begin{equation}
  \mathcal{H}=\hat{\mathcal{A}}_d^\dagger\hat{\mathcal{A}}_d
  +\tilde{\mathcal{E}}_d,\quad
  \hat{\mathcal{A}}_d\eqdef\frac{d}{dx}
  -\frac{\partial_x\tilde{\phi}_d(x)}{\tilde{\phi}_d(x)},\quad
  \forall d\in\mathcal{V}.
\end{equation}
Since $\tilde{\phi}_d(x)$ has no node, the operator $\hat{\mathcal{A}}_d$
and $\hat{\mathcal{A}}_d^\dagger$ are regular in the interior.
Next we define $\mathcal{H}^{[1]}$ by interchanging the order of these
two operators:
\begin{equation}
  \mathcal{H}^{[1]}\eqdef\hat{\mathcal{A}}_d\hat{\mathcal{A}}_d^\dagger
  +\tilde{\mathcal{E}}_d,
  \ \ U^{[1]}(x)=U(x)-2\partial_x^2\log\bigl|\tilde{\phi}_d(x)\bigr|.
\end{equation}
It is straightforward to show that $\mathcal{H}$ and $\mathcal{H}^{[1]}$
are iso-spectral
\begin{align*}
  &\mathcal{H}^{[1]}\phi_n^{[1]}(x)=\mathcal{E}_n\phi_n^{[1]}(x),
%  \label{philform1}
  \\
  &\phi_n^{[1]}(x)\eqdef\hat{\mathcal{A}}_d\phi_n(x)
  =\frac{\text{W}[\tilde{\phi}_d,\phi_n](x)}{\tilde{\phi}_d(x)}\quad
  (n\in\mathbb{Z}_{\ge0}),
%  \label{philform2}
  \\
  &\mathcal{H}^{[1]}\tilde{\phi}_{\text{v}}^{[1]}(x)
  =\tilde{\mathcal{E}}_{\text{v}}\tilde{\phi}_{\text{v}}^{[1]}(x),\\
  &\tilde{\phi}_{\text{v}}^{[1]}(x)\eqdef
  \hat{\mathcal{A}}_d\tilde{\phi}_{\text{v}}(x)
  =\frac{\text{W}[\tilde{\phi}_d,\tilde{\phi}_{\text{v}}](x)}
  {\tilde{\phi}_d(x)}\quad
  (\text{v}\in\mathcal{V}\backslash\{d\}),\\
  &\phi_n(x)=\frac{\hat{\mathcal{A}}_d^\dagger}
  {\mathcal{E}_n-\tilde{\mathcal{E}}_d}\phi_n^{[1]}(x),\\
  &(\phi_m^{[1]},\phi_n^{[1]})=(\mathcal{E}_n-\tilde{\mathcal{E}}_d)
  (\phi_m,\phi_n)=(\mathcal{E}_n-\tilde{\mathcal{E}}_d)h_n\delta_{m\,n}.
%  \label{ulform}
\end{align*}
Obviously the  state $d$ is {\em deleted} from the virtual state spectrum
of $\mathcal{H}^{[1]}$, for $\hat{\mathcal{A}}_d\tilde{\phi}_d(x)=0$.
Since the zero-modes of $\hat{\mathcal{A}}_d$ and
$\hat{\mathcal{A}}_d^\dagger$ are $\tilde{\phi}_d(x)$ and
$1/\tilde{\phi}_d(x)$, which have infinite norms, the sets of
eigenfunctions $\{\phi_n(x)\}$ and $\{\phi_n^{[1]}(x)\}$ are iso-spectral
and in one to one correspondence.
Thus the new set $\{\phi_n^{[1]}(x)\}$ is {\em complete}. If not,
there must exist a finite norm vector $\Psi$ in the Hilbert space which
is orthogonal to all the basis:
$(\Psi,\phi_n^{[1]})=0$, $\forall n\in\mathbb{Z}_{\ge0}$. This would lead
to a contradiction $(\hat{\mathcal{A}}_d^\dagger\Psi,\phi_n)=0$,
$\forall n\in\mathbb{Z}_{\ge0}$, since $\hat{\mathcal{A}}_d^\dagger\Psi\neq0$.
Next we show the logic for demonstrating that the virtual state solutions
$\{\tilde{\phi}_{\text{v}}^{[1]}(x)\}$ have no node in the interior.
By using the Schr\"{o}dinger equations for them we obtain
\begin{equation}
  \partial_x\text{W}[\tilde{\phi}_d,\tilde{\phi}_{\text{v}}](x)
  =\bigl(\tilde{\mathcal{E}}_d-\tilde{\mathcal{E}}_{\text{v}}\bigr)
  \tilde{\phi}_d(x)\tilde{\phi}_{\text{v}}(x),
\end{equation}
which has no node. Since  we can verify that
$\text{W}[\tilde{\phi}_d,\tilde{\phi}_{\text{v}}](x)$ vanishes at one
boundary, no-nodeness of $\text{W}[\tilde{\phi}_d,\tilde{\phi}_{\text{v}}](x)$
in the interior follows.
When the virtual states $d$ and $\text{v}$ belong to different types
(I and II) the Wronskians might not vanish at both boundaries.
In that case we can show that they have the same sign at both boundaries
$W[\tilde{\phi}_d,\tilde{\phi}_{\text{v}}](x_1)
W[\tilde{\phi}_d,\tilde{\phi}_{\text{v}}](x_2)>0$.
Likewise we can show, for all the explicit examples in the next section
that $\tilde{\phi}_\text{v}^{[1]}$ and $1/\tilde{\phi}_\text{v}^{[1]}$
have infinite norms.
The steps going from $\mathcal{H}^{[1]}\to \mathcal{H}^{[2]}$ and further
are essentially the same.

%%%%%%%%%%%%%%%%%%%%%%%%%%%%%%%%%%%%%%%%%%%%%%%%%%%%%%%%%%%%%%%
%                                                             %
%  3. Multi-indexed Laguerre and Jacobi Polynomials           %
%                                                             %
%%%%%%%%%%%%%%%%%%%%%%%%%%%%%%%%%%%%%%%%%%%%%%%%%%%%%%%%%%%%%%%
\section{Multi-indexed Laguerre and Jacobi Polynomials}
\label{sec:Multi}

Here we will apply the method of virtual states deletion to two
well-known systems of exactly solvable quantum mechanics with the
radial oscillator and  Darboux-P\"oschl-Teller potentials:
\begin{equation}
  U(x)=\left\{
  \begin{array}{ll}
  {\displaystyle x^2+\frac{g(g-1)}{x^2}-(1+2g)},&\\
  \qquad x_1=0,\ x_2=\infty,\
  g>\frac12,&:\text{L}\\[5pt]
  {\displaystyle \frac{g(g-1)}{\sin^2x}+\frac{h(h-1)}{\cos^2x}-(g+h)^2},&\\[4pt]
   \qquad x_1=0,\ x_2=\frac{\pi}{2},\ g,\,h>\frac12,&:\text{J}
  \end{array}\right.,
\end{equation}
in which L and J stand for the names of their eigenfunctions, the Laguerre
and Jacobi polynomials. For the actual parameter ranges consistent with
specific deletions, see \eqref{Lbound}-\eqref{Jbound}.
The eigenfunctions are factorised into the groundstate eigenfunction
and the polynomial in the {\em sinusoidal coordinate} $\eta=\eta(x)$
\cite{os7}:
\begin{equation}
  \phi_n(x;\bm{\lambda})=\phi_0(x;\bm{\lambda})P_n(\eta(x);\bm{\lambda}),
  \label{factor}
\end{equation}
in which $\bm{\lambda}$ stands for the parameters, $g$ for L and $(g,h)$
for J. Their explicit forms are
\begin{align}
  \text{L}:&\ \ \phi_0(x;g)\eqdef e^{-\frac12x^2}x^g,\quad\eta(x)\eqdef x^2,\n
  &\ \ P_n(\eta;g)\eqdef L_n^{(g-\frac12)}(\eta),\n
  &\ \ \mathcal{E}_n(g)\eqdef 4n,\quad
  h_n(g)\eqdef\frac{1}{2n!}\Gamma(n+g+\tfrac12),
  \label{phi0L} \\
  \text{J}:&\ \ \phi_0(x;g,h)\eqdef(\sin x)^g(\cos x)^h,\quad
  \eta(x)\eqdef\cos2x,\n
  &\ \ P_n(\eta;g,h)\eqdef P_n^{(g-\frac12,h-\frac12)}(\eta),\n
  &\ \ \mathcal{E}_n(g,h)\eqdef 4n(n+g+h),\n
  &\ \ h_n(g,h)\eqdef\frac{\Gamma(n+g+\frac12)\Gamma(n+h+\frac12)}
  {2n!(2n+g+h)\Gamma(n+g+h)}.
  \label{phi0J}
\end{align}

They have two types of virtual states, which are again {\em polynomial
solutions}. The total number is finite depending on the parameter values
$g$ and $h$, except for L1 which has infinite.
The virtual states wavefunctions for L are:
\begin{align}
 \text{L1}:&\ \ \tilde{\phi}_\text{v}^\ai(x)\eqdef
  e^{\frac12x^2}x^g\xi_\text{v}^\ai(\eta(x);g),\n
  &\ \ \xi_\text{v}^\ai(\eta;g)\eqdef P_\text{v}(-\eta;g),
  \quad\text{v}\in\mathbb{Z}_{\ge0},\n
  &\ \ \tilde{\mathcal{E}}_\text{v}^\ai\eqdef-4(g+\text{v}+\tfrac12),\quad
  \tilde{\bm{\delta}}^{\ai}\eqdef-1,
  \label{vsL1}\\
  \text{L2}:&\ \ \tilde{\phi}_\text{v}^\ait(x)\eqdef
  e^{-\frac12x^2}x^{1-g}\xi_\text{v}^\ait(\eta(x);g),\n
  &\ \ \xi_\text{v}^\ait(\eta;g)\eqdef P_\text{v}(\eta;1-g),
  \quad\text{v}=0,1,\ldots,[g-\tfrac12]',\n
  &\ \ \tilde{\mathcal{E}}_\text{v}^\ait\eqdef-4(g-\text{v}-\tfrac12),\quad
  \tilde{\bm{\delta}}^{\ait}\eqdef 1,
  \label{vsL2}
\end{align}
in which $[a]'$ denotes the greatest integer less than $a$ and
$\tilde{\bm{\delta}}^{\ai,\ait}$ will be used later.
For no-nodeness of $\xi_{\text{v}}^{\ai,\ait}$, see (2.39) of \cite{os18}.
The virtual states wavefunctions for J are:
\begin{align}
  \!\!\text{J1}:&\ \tilde{\phi}_\text{v}^\ai(x)\eqdef(\sin x)^g(\cos x)^{1-h}
  \xi_\text{v}^\ai(\eta(x);g,h),\n
  \!\!&\ \xi_\text{v}^\ai(\eta;g,h)\eqdef P_\text{v}(\eta;g,1-h),
  \ \ \text{v}=0,1,\ldots,[h-\tfrac12]',\n
  \!\!&\ \tilde{\mathcal{E}}_\text{v}^\ai\eqdef-4(g+\text{v}+\tfrac12)
  (h-\text{v}-\tfrac12),
  \ \ \tilde{\bm{\delta}}^{\ai}\eqdef(-1,1),
  \label{vsJ1}\\
  \!\!\text{J2}:&\ \tilde{\phi}_\text{v}^\ait(x)\eqdef(\sin x)^{1-g}(\cos x)^h
  \xi_\text{v}^\ait(\eta(x);g,h),\n
  \!\!&\ \xi_\text{v}^\ait(\eta;g,h)\eqdef P_\text{v}(\eta;1-g,h),
  \ \ \text{v}=0,1,\ldots,[g-\tfrac12]',\n
  \!\!&\ \tilde{\mathcal{E}}_\text{v}^\ait\eqdef-4(g-\text{v}-\tfrac12)
  (h+\text{v}+\tfrac12),
  \ \,\tilde{\bm{\delta}}^{\ait}\eqdef(1,-1).\!\!
  \label{vsJ2}
\end{align}
The larger the parameter $g$ and/or $h$ become, the more the virtual states
are `created'. 
The L1 system is obtained from the J1 in the limit $h\to\infty$ \cite{os19}.
This explains the infinitely many virtual states of L1.
Let us denote by $\mathcal{V}^{\ai,\ait}$ the index sets of
the virtual states of type I and II for L and J.
Due to the parity property of the Jacobi polynomial
$P_n^{(\alpha,\beta)}(-x)=(-1)^nP_n^{(\beta,\alpha)}(x)$, the two virtual
state polynomials $\xi_\text{v}^\ai$ and $\xi_\text{v}^\ait$ for J are
related by
$\xi_\text{v}^\ait(-\eta;g,h)=(-1)^\text{v}\xi_\text{v}^\ai(\eta;h,g)$.
For no-nodeness of $\xi_{\text{v}}^{\ai,\ait}$, see (2.40) of \cite{os18}
and (3.2) of \cite{os21}.
It is also obvious that the virtual state wavefunctions
$\tilde{\phi}_\text{v}$ and their reciprocals $1/\tilde{\phi}_\text{v}$
are not square integrable in all four types.
There is a third type of negative energy polynomial solutions for both L
and J. For J, the degree $n$ solution is
$(\sin x)^{1-g}(\cos x)^{1-h}P_n^{(-g+\frac12,-h+\frac12)}(\eta)$.
Even when they have no node in the interior, their reciprocals are square
integrable and they cannot be used for the virtual states deletion.
The label 0 is special in that the wavefunctions satisfy
$\tilde{\phi}_0^\ai(x;\bm{\lambda})\tilde{\phi}_0^\ait(x;\bm{\lambda})=
\cF^{-1}\frac{d\eta(x)}{dx}$ since $\xi_0^{\ai,\ait}=1$.
Here the constant $\cF=2$ for L and $\cF=-4$ for J.
We will not use the label 0 states for deletion.

%%%%%%%%%%%%%%%%%%%%%%%%%%%%%%%%%%%%%%%%%%%%%%%%%%%%%%%%%%%%%%%%%%%%%%%%
\begin{center}
  \includegraphics[scale=0.8]{virtualadler2.epsi}
\end{center}
\begin{center}
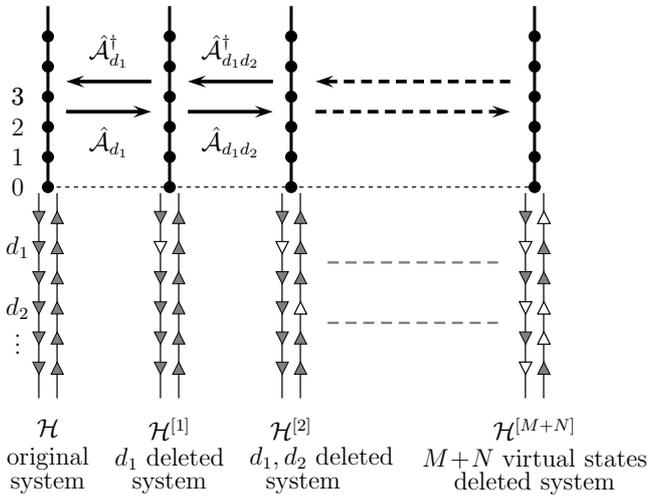
Figure\,1: Schematic picture of virtual states deletion.\\
The black circles denote eigenstates.
The down and up triangles denote virtual states of type I and II.
The deleted virtual states are denoted by white triangles.
\end{center}
%%%%%%%%%%%%%%%%%%%%%%%%%%%%%%%%%%%%%%%%%%%%%%%%%%%%%%%%%%%%%%%%%%%%%%%%

Since there are two types of virtual states available, the general deletion
is specified by the set of $M+N$ positive integers
$\mathcal{D}\eqdef\{d_1^\ai,\ldots,d_M^\ai,d_1^\ait,\ldots,d_N^\ait\}$,
$d_j^{\ai,\ait}\ge1$, which are the degrees of the deleted virtual state
wavefunctions.
Note that the subcases of either $M=0$ or $N=0$ are meaningful.
In order to accommodate all these virtual states, the parameters $g$ and
$h$ must be larger than certain bounds:
\begin{align}
  \text{L}:&\ \ g>\text{max}\{N+\tfrac32,d_j^\ait+\tfrac12\},
  \label{Lbound}\\
  \text{J}:&\ \ g>\text{max}\{N+2,d_j^\ait+\tfrac12\},\n
  &\ \ h>\text{max}\{M+2,d_j^\ai+\tfrac12\}.
  \label{Jbound}
\end{align}
See Figure 1 for the schematic structure of virtual states deletion.

Like the original eigenfunctions \eqref{factor} the $n$-th eigenfunction
$\phi_{\mathcal{D},n}(x;\bm{\lambda})\equiv\phi_n^{[M,N]}(x)$ after the
deletion \eqref{fineig} can be clearly factorised into an $x$-dependent
part, the common denominator polynomial $\Xi_{\mathcal{D}}$ in $\eta$
and the {\em multi-indexed polynomial} $P_{\mathcal{D},n}$ in $\eta$:
\begin{align}
  \phi_n^{[M,N]}(x)\equiv\phi_{\mathcal{D},n}(x;\bm{\lambda})
  &=\cF^{M+N}\psi_{\mathcal{D}}(x;\bm{\lambda})
  P_{\mathcal{D},n}(\eta(x);\bm{\lambda}),\n
  \psi_{\mathcal{D}}(x;\bm{\lambda})&\eqdef
  \frac{\phi_0(x;\bm{\lambda}^{[M,N]})}
  {\Xi_{\mathcal{D}}(\eta(x);\bm{\lambda})},
\end{align}
in which $\phi_0(x;\bm{\lambda})$ is the groundstate wavefunction
\eqref{phi0L}-\eqref{phi0J}.
Here the shifted parameters $\bm{\lambda}^{[M,N]}$ after the $[M,N]$ deletion
are $\bm{\lambda}^{[M,N]}\eqdef\bm{\lambda}-M\tilde{\bm{\delta}}^\ai
-N\tilde{\bm{\delta}}^\ait$, explicitly they are
\begin{equation}
  \bm{\lambda}^{[M,N]}=\left\{
  \begin{array}{ll}
  g+M-N&:\text{L}\\
  (g+M-N,h-M+N)&:\text{J}
  \end{array}\right..
\end{equation}
Needless to say, the denominator polynomial $\Xi_{\mathcal{D}}$ has no
node in the interior.
The polynomials $P_{\mathcal{D},n}$ and $\Xi_{\mathcal{D}}$ are expressed
in terms of Wronskians of the variable $\eta$:
\begin{align}
  P_{\mathcal{D},n}(\eta;\bm{\lambda})&\eqdef
  \text{W}[\mu_1,\ldots,\mu_M,\nu_1,\ldots,\nu_N,P_n](\eta)\n
  &\hspace*{-8mm}\times\left\{
  \begin{array}{ll}
  e^{-M\eta}\,\eta^{(M+g+\frac12)N}&:\text{L}\\[2pt]
  \bigl(\frac{1-\eta}{2}\bigr)^{(M+g+\frac12)N}
  \bigl(\frac{1+\eta}{2}\bigr)^{(N+h+\frac12)M}&:\text{J}
  \end{array}\right.,
  \label{multiP}\\
  \Xi_{\mathcal{D}}(\eta;\bm{\lambda})&\eqdef
  \text{W}[\mu_1,\ldots,\mu_M,\nu_1,\ldots,\nu_N](\eta)\n
  &\hspace*{-8mm}\times\left\{
  \begin{array}{ll}
  e^{-M\eta}\,\eta^{(M+g-\frac12)N}&:\text{L}\\[2pt]
  \bigl(\frac{1-\eta}{2}\bigr)^{(M+g-\frac12)N}
  \bigl(\frac{1+\eta}{2}\bigr)^{(N+h-\frac12)M}&:\text{J}
  \end{array}\right.,\\
  &\ \ \,\quad
  \mu_j=\left\{
  \begin{array}{ll}
  e^{\eta}\xi_{d_j^\ai}^\ai(\eta;g)&:\text{L}\\[4pt]
  \bigl(\frac{1+\eta}{2}\bigr)^{\frac12-h}\xi_{d_j^\ai}^\ai(\eta;g,h)
  &:\text{J}
  \end{array}\right.,\n
  &\ \ \,\quad
  \nu_j=\left\{
  \begin{array}{ll}
  \eta^{\frac12-g}\xi_{d_j^\ait}^\ait(\eta;g)&:\text{L}\\[4pt]
  \bigl(\frac{1-\eta}{2}\bigr)^{\frac12-g}\xi_{d_j^\ait}^\ait(\eta;g,h)
  &:\text{J}
  \end{array}\right.,
\end{align}
in which $P_n$ in \eqref{multiP} denotes the original polynomial,
$P_n(\eta;g)$ for L and $P_n(\eta;g,h)$ for J.
These polynomials depend on the order of deletions only through
the sign of permutation.
The multi-indexed polynomial $P_{\mathcal{D},n}$ is of degree $\ell+n$
and the denominator polynomial $\Xi_{\mathcal D}$ is of degree $\ell$ in
$\eta$, in which $\ell$ is given by
\begin{equation}
  \ell\eqdef\sum_{j=1}^Md_j^\ai+\sum_{j=1}^Nd_j^\ait
  -\frac12 M(M-1)-\frac12N(N-1)+MN\ge1.
\end{equation}
Here the label $n$ specifies the energy eigenvalue $\mathcal{E}_n$ of
$\phi_{\mathcal{D},n}$. Hence it also counts the nodes due to the
oscillation theorem.
Although they do not satisfy the three term recurrence relations, the
multi-indexed polynomials $\{P_{\mathcal{D},n}\}$ form a complete set of
orthogonal polynomials with the orthogonality relations:
\begin{align}
  &\quad\int\!\!d\eta\,
  \frac{W(\eta;\bm{\lambda}^{[M,N]})}
  {\Xi_{\mathcal{D}}(\eta;\bm{\lambda})^2}
  P_{\mathcal{D},m}(\eta;\bm{\lambda})
  P_{\mathcal{D},n}(\eta;\bm{\lambda})\n
  &=h_n(\bm{\lambda})\delta_{nm}\n
  &\ \times\!\left\{
  \begin{array}{ll}
  \prod_{j=1}^M(n+g+d_j^\ai+\tfrac12)\cdot
  \prod_{j=1}^N(n+g-d_j^\ait-\tfrac12)
  &\!\!\!\!\!:\text{L}\\[3pt]
  4^{-M-N}
  \prod_{j=1}^M(n+g+d_j^\ai+\tfrac12)(n+h-d_j^\ai-\tfrac12)\\[2pt]
  \qquad\times
  \prod_{j=1}^N(n+g-d_j^\ait-\tfrac12)(n+h+d_j^\ait+\tfrac12)
  &\!\!\!\!\!:\text{J}
  \end{array}\right.\!\!\!,
\end{align}
where the weight function of the original polynomials
$W(\eta;\bm{\lambda})d\eta=\phi_0(x;\bm{\lambda})^2dx$ reads explicitly
\begin{equation}
  W(\eta;\bm{\lambda})\eqdef
  \left\{\begin{array}{ll}
  \frac12e^{-\eta}\eta^{g-\frac12}&:\text{L}\\
  \frac{1}{2^{g+h+1}}(1-\eta)^{g-\frac12}(1+\eta)^{h-\frac12}&:\text{J}
  \end{array}\right..
\end{equation}
This is obtained by rewriting the orthogonality relations of the
eigenfunctions \eqref{fineig} after the $[M,N]$ deletion.

In the rest of this section we explore various properties of the new
multi-indexed polynomials $\{P_{\mathcal{D},n}\}$. As for the exceptional
orthogonal polynomials \cite{os16,os17,os19,hos,os20,os23}, the lowest
degree polynomial $P_{\mathcal{D},0}(\eta;\bm{\lambda})$ is related to
the denominator polynomial $\Xi_{\mathcal D}(\eta;\bm{\lambda})$ by the
parameter shift $\bm{\lambda}\to\bm{\lambda}+\bm{\delta}$ ($\bm{\delta}=1$
for L and $\bm{\delta}=(1,1)$ for J):
\begin{align}
  &P_{\mathcal{D},0}(\eta;\bm{\lambda})
  =\Xi_{\mathcal{D}}(\eta;\bm{\lambda}+\bm{\delta})\n
  &\hspace{8mm}\times\left\{
  \begin{array}{ll}
  (-1)^M\prod_{j=1}^N(g-d_j^\ait-\tfrac12)&:\text{L}\\[4pt]
  2^{-M}\prod_{j=1}^M(h-d_j^\ai-\tfrac12)&\\[2pt]
  \ \ \times(-2)^{-N}\prod_{j=1}^{N}(g-d_j^\ait-\tfrac12)&:\text{J}
  \end{array}\right.\!.
  \label{plusdelta}
\end{align}
The virtual state wavefunction with the index $\text{v}\in{\mathcal D}$
leading to the final $[M,N]$ deletion is $\tilde{\phi}_\text{v}^{[M',N']}$
\eqref{finvirt} with the index set
$\mathcal{D}'\eqdef\mathcal{D}\backslash\{\text{v}\}$, in which
$[M',N']=[M-1,N]$ if $\text{v}\in\mathcal{V}^\ai$, and $[M,N-1]$ if
$\text{v}\in\mathcal{V}^\ait$. It has the following form:
\begin{align}
& \tilde{\phi}^{[M',N']}_{\text{v}}(x;\bm{\lambda})
  =\cF^{M+N-1}
  \frac{\Xi_{\mathcal{D}}(\eta(x);\bm{\lambda})}
  {\Xi_{\mathcal{D}'}(\eta(x);\bm{\lambda})}\n
  &\times\left\{
  \begin{array}{ll}
  (-1)^N\phi_0(ix;\bm{\lambda}^{[M-1,N]})i^{-(g+M-N-1)},
  &\!\!\!\text{v}\in\mathcal{V}^\ai\ :\text{L}\\[2pt]
  \phi_0(x;\mathfrak{t}^\ait(\bm{\lambda}^{[M,N-1]})),
  &\!\!\!\text{v}\in\mathcal{V}^\ait :\text{L}\\[2pt]
  (-1)^N\phi_0(x;\mathfrak{t}^\ai(\bm{\lambda}^{[M-1,N]})),
  &\!\!\!\text{v}\in\mathcal{V}^\ai\ :\text{J}\\[2pt]
  \phi_0(x;\mathfrak{t}^\ait(\bm{\lambda}^{[M,N-1]})),
  &\!\!\!\text{v}\in\mathcal{V}^\ait :\text{J}
  \end{array}\right.\!\!.
\end{align}
Here the twist operator $\mathfrak{t}$ acting on the parameters is
defined as $\mathfrak{t}^\ait(g)=1-g$ for L,
$\mathfrak{t}^\ai(g,h)=(g,1-h)$ and $\mathfrak{t}^\ait(g,h)=(1-g,h)$ for J.
Note that the virtual states \eqref{vsL2}--\eqref{vsJ2} have the form
$\tilde{\phi}_{\text{v}}(x;\bm{\lambda})
=\phi_{\text{v}}(x;\mathfrak{t}(\bm{\lambda}))$.
These relations are essential for the determination of the lower bounds
of the parameters given in \eqref{Lbound}--\eqref{Jbound}, which guarantee
that the virtual state properties are correctly inherited at each step
of deletion.

The Hamiltonian
$\mathcal{H}_{\mathcal D}(\bm{\lambda})\equiv\mathcal{H}^{[M,N]}$ of
the $[M,N]$ deleted system can be expressed in terms of its groundstate
eigenfunction $\phi_{\mathcal{D},0}(x;\bm{\lambda})\equiv
\phi_0^{[M,N]}(x;\bm{\lambda})$ with the help of
\eqref{plusdelta}:
\begin{align}
  &\mathcal{H}_{\mathcal D}(\bm{\lambda})
  =\mathcal{A}_{\mathcal D}(\bm{\lambda})^\dagger
  \mathcal{A}_{\mathcal D}(\bm{\lambda}),
  \label{hamD}\\
  &\mathcal{A}_{\mathcal D}(\bm{\lambda})\eqdef
  \frac{d}{dx}-\frac{\partial_x\phi_{\mathcal{D},0}(x;\bm{\lambda})}
  {\phi_{\mathcal{D},0}(x;\bm{\lambda})},\n
  &\phi_{\mathcal{D},0}(x;\bm{\lambda})\propto
  \phi_{0}(x;\bm{\lambda}^{[M,N]})
  \frac{\Xi_{\mathcal D}(\eta(x);\bm{\lambda}+\bm{\delta})}
  {\Xi_{\mathcal D}(\eta(x);\bm{\lambda})}.
  \label{phiD0}
\end{align}
This has the same form as the various Hamiltonians of the exceptional
orthogonal polynomials including those of the discrete quantum mechanics
\cite{os16,os17,os19,os23}.
Reflecting the construction \cite{stz,os21} it is shape invariant
\cite{genden,os16,os18}
\begin{equation}
  \mathcal{A}_{\mathcal{D}}(\bm{\lambda})
  \mathcal{A}_{\mathcal{D}}(\bm{\lambda})^{\dagger}
  =\mathcal{A}_{\mathcal{D}}(\bm{\lambda}+\bm{\delta})^{\dagger}
  \mathcal{A}_{\mathcal{D}}(\bm{\lambda}+\bm{\delta})
  +\mathcal{E}_1(\bm{\lambda}).
  \label{shapeinvD}
\end{equation}
This means that the operators $\mathcal{A}_{\mathcal{D}}(\bm{\lambda})$ and
$\mathcal{A}_{\mathcal{D}}(\bm{\lambda})^\dagger$ relate the eigenfunctions
of neighbouring degrees and parameters:
\begin{align}
  \mathcal{A}_{\mathcal{D}}(\bm{\lambda})
  \phi_{\mathcal{D},n}(x;\bm{\lambda})
  &=f_n(\bm{\lambda})
  \phi_{\mathcal{D},n-1}(x;\bm{\lambda}+\bm{\delta}),
  \label{ADphiDn=}\\
  \mathcal{A}_{\mathcal{D}}(\bm{\lambda})^{\dagger}
  \phi_{\mathcal{D},n-1}(x;\bm{\lambda}+\bm{\delta})
  &=b_{n-1}(\bm{\lambda})
  \phi_{\mathcal{D},n}(x;\bm{\lambda}),
  \label{ADdphiDn=}
\end{align}
in which the constants  $f_n(\bm{\lambda})$ and $b_{n-1}(\bm{\lambda})$
are the factors of the eigenvalue
$f_n(\bm{\lambda})b_{n-1}(\bm{\lambda})=\mathcal{E}_n(\bm{\lambda})$:
\begin{equation*}
  f_n(\bm{\lambda})=\left\{
  \begin{array}{ll}
  -2&:\text{L}\\
  -2(n+g+h)&:\text{J}
  \end{array}\right.\!\!\!,
  \ b_{n-1}(\bm{\lambda})=-2n\ \ :\text{L,\,J}.
\end{equation*}
The forward and backward shift operators are defined by
\begin{align}
  \mathcal{F}_{\mathcal{D}}(\bm{\lambda})&\eqdef
  \psi_{\mathcal{D}}\,(x;\bm{\lambda}+\bm{\delta})^{-1}\circ
  \mathcal{A}_{\mathcal{D}}(\bm{\lambda})\circ
  \psi_{\mathcal{D}}\,(x;\bm{\lambda})
  \label{FDdef}\\
  &=\cF\frac{\Xi_{\mathcal{D}}(\eta;\bm{\lambda}+\bm{\delta})}
  {\Xi_{\mathcal{D}}(\eta;\bm{\lambda})}\Bigl(\frac{d}{d\eta}
  -\frac{\partial_{\eta}\Xi_{\mathcal{D}}(\eta;\bm{\lambda}+\bm{\delta})}
  {\Xi_{\mathcal{D}}(\eta;\bm{\lambda}+\bm{\delta})}\Bigr),
  \label{FD}\\
  \mathcal{B}_{\mathcal{D}}(\bm{\lambda})&\eqdef
  \psi_{\mathcal{D}}\,(x;\bm{\lambda})^{-1}\circ
  \mathcal{A}_{\mathcal{D}}(\bm{\lambda})^{\dagger}\circ
  \psi_{\mathcal{D}}\,(x;\bm{\lambda}+\bm{\delta})
  \label{BDdef}\\
  &=-4\cF^{-1}c_2(\eta)\frac{\Xi_{\mathcal{D}}(\eta;\bm{\lambda})}
  {\Xi_{\mathcal{D}}(\eta;\bm{\lambda}+\bm{\delta})}\n
  &\qquad\times\Bigl(\frac{d}{d\eta}
  +\frac{c_1(\eta,\bm{\lambda}^{[M,N]})}{c_2(\eta)}
  -\frac{\partial_{\eta}\Xi_{\mathcal{D}}(\eta;\bm{\lambda})}
  {\Xi_{\mathcal{D}}(\eta;\bm{\lambda})}\Bigr),
  \label{BD}
\end{align}
in which the functions $c_1(\eta;\bm{\lambda})$ and $c_2(\eta)$ are those
appearing in the (confluent) hypergeometric equations for the Laguerre and
Jacobi polynomials
\begin{align}
  \text{L}:&\ \ c_1(\eta,\bm{\lambda})\eqdef g+\tfrac12-\eta,\quad
  c_2(\eta)\eqdef\eta,\n
  \text{J}:&\ \ c_1(\eta,\bm{\lambda})\eqdef h-g-(g+h+1)\eta,\quad
  c_2(\eta)\eqdef 1-\eta^2.
  \nonumber
\end{align}
Their action on the multi-indexed polynomials
$P_{\mathcal{D},n}(\eta;\bm{\lambda})$ is
\begin{align}
  \mathcal{F}_{\mathcal{D}}(\bm{\lambda})
  P_{\mathcal{D},n}(\eta;\bm{\lambda})
  &=f_n(\bm{\lambda})
  P_{\mathcal{D},n-1}(\eta;\bm{\lambda}+\bm{\delta}),
  \label{FDPDn=}\\
  \mathcal{B}_{\mathcal{D}}(\bm{\lambda})
  P_{\mathcal{D},n-1}(\eta;\bm{\lambda}+\bm{\delta})
  &=b_{n-1}(\bm{\lambda})
  P_{\mathcal{D},n}(\eta;\bm{\lambda}).
  \label{BDPDn=}
\end{align}
The second order differential operator
$\widetilde{\mathcal{H}}_{\mathcal{D}}(\bm{\lambda})$ governing the
multi-indexed polynomials is:
\begin{align}
  &\widetilde{\mathcal{H}}_{\mathcal{D}}(\bm{\lambda})
  \eqdef\psi_{\mathcal{D}}(x;\bm{\lambda})^{-1}\circ
  \mathcal{H}_{\mathcal{D}}(\bm{\lambda})\circ
  \psi_{\mathcal{D}}(x;\bm{\lambda})
  =\mathcal{B}_{\mathcal{D}}(\bm{\lambda})
  \mathcal{F}_{\mathcal{D}}(\bm{\lambda})\n
  &=-4\biggl(c_2(\eta)\frac{d^2}{d\eta^2}\!
  +\!\Bigl(c_1(\eta,\bm{\lambda}^{[M,N]})-2c_2(\eta)
  \frac{\partial_{\eta}\Xi_{\mathcal{D}}(\eta;\bm{\lambda})}
  {\Xi_{\mathcal{D}}(\eta;\bm{\lambda})}\Bigr)\frac{d}{d\eta}\n
  &\qquad
  +c_2(\eta)
  \frac{\partial^2_{\eta}\Xi_{\mathcal{D}}(\eta;\bm{\lambda})}
  {\Xi_{\mathcal{D}}(\eta;\bm{\lambda})}
  -c_1(\eta,\bm{\lambda}^{[M,N]}-\bm{\delta})
  \frac{\partial_{\eta}\Xi_{\mathcal{D}}(\eta;\bm{\lambda})}
  {\Xi_{\mathcal{D}}(\eta;\bm{\lambda})}
  \biggr),
  \label{ThamD}\\
  &\widetilde{\mathcal{H}}_{\mathcal{D}}(\bm{\lambda})
  P_{\mathcal{D},n}(\eta;\bm{\lambda})=\mathcal{E}_n(\bm{\lambda})
  P_{\mathcal{D},n}(\eta;\bm{\lambda}).
  \label{fuchs}
\end{align}
Since all the zeros of $\Xi_{\mathcal{D}}(\eta;\bm{\lambda})$ are simple,
\eqref{fuchs} is a Fuchsian differential equation for the J case.
The characteristic exponents at the zeros of
$\Xi_{\mathcal{D}}(\eta;\bm{\lambda})$ are the same everywhere, 0 and 3.
The multi-indexed polynomials $\{P_{\mathcal{D},n}(\eta;\bm{\lambda})\}$
provide infinitely many global solutions of the above Fuchsian equation
\eqref{fuchs} with $3+\ell$ regular singularities for the J case
\cite{hos}. The L case is obtained in a confluent limit. These situations
are basically the same as those of the exceptional polynomials.

Although we have restricted $d_j^{\ai,\ait}\geq 1$, there is no
obstruction for deletion of $d_j^{\ai,\ait}=0$.
In terms of the multi-indexed polynomial \eqref{multiP}, the level 0
deletions imply the following:
\begin{align}
  &P_{\mathcal{D},n}(\eta;\bm{\lambda})\Bigm|_{d_M^{\ai}=0}
  =P_{\mathcal{D}',n}(\eta;\bm{\lambda}-\tilde{\bm{\delta}}^{\ai})
  \times A,\n
  &\quad
  \mathcal{D}'=\{d_1^{\ai}-1,\ldots,d_{M-1}^{\ai}-1,
  d_1^{\ait}+1,\ldots,d_N^{\ait}+1\},
  \label{dIM=0}\\
  &P_{\mathcal{D},n}(\eta;\bm{\lambda})\Bigm|_{d_N^{\ait}=0}
  =P_{\mathcal{D}',n}(\eta;\bm{\lambda}-\tilde{\bm{\delta}}^{\ait})
  \times B,\n
  &\quad
  \mathcal{D}'=\{d_1^{\ai}+1,\ldots,d_M^{\ai}+1,
  d_1^{\ait}-1,\ldots,d_{N-1}^{\ait}-1\},
  \label{dIIN=0}
\end{align}
where the multiplicative factors $A$ and $B$ are
\begin{align*}
  A&=\left\{
  \begin{array}{ll}
  (-1)^M\prod_{j=1}^N(d_j^{\ait}+1)&:\text{L}\\[4pt]
  -(-2)^{-M}\prod_{j=1}^{M-1}(g-h+d_j^{\ai}+1)&\\[3pt]
  \ \ \times(-2)^{-N}\prod_{j=1}^N(d_j^{\ait}+1)\cdot
  (n+h-\frac12)&:\text{J}
  \end{array}\right.\!\!\!,\\
  B&=\left\{
  \begin{array}{ll}
  (-1)^M\prod_{j=1}^M(d_j^{\ai}+1)\cdot
  (n+g-\frac12)&:\text{L}\\[4pt]
  2^{-M}\prod_{j=1}^M(d_j^{\ai}+1)\cdot(-2)^{-N}&\\[3pt]
  \ \times\prod_{j=1}^{N-1}(h-g+d_j^{\ait}+1)\cdot
  (n+g-\frac12)&:\text{J}
  \end{array}\right.\!\!\!.
\end{align*}
{}From \eqref{plusdelta}, $\Xi_{\mathcal{D}}$ behaves similarly.
Therefore including the level 0 deletion corresponds to $M+N-1$ virtual states
deletions. This is why we have restricted $d_j^{\ai,\ait}\ge1$.

These relations \eqref{dIM=0}--\eqref{dIIN=0} can be used for studying
the equivalence of $\mathcal{H}_{\mathcal{D}}$.
It should be stressed that the same polynomials can have different but
equivalent sets of multi-indices $\mathcal{D}$, which defines the
denominator polynomial $\Xi_{\mathcal D}$.
Since the overall scale of $\Xi_{\mathcal D}$ is irrelevant (see
\eqref{phiD0}) for the multi-indexed polynomial $\{P_{\mathcal{D},n}\}$,
we consider the equivalence class of $\Xi_{\mathcal D}$.
For example, it is straightforward to verify the following equivalence
and their duals $\ai\leftrightarrow\ait$:
\begin{align}
  &\Xi_{\{1^\ai,\ldots,k^\ai\}}(\eta;\bm{\lambda}-\tilde{\bm{\delta}}^\ait)
  \sim\Xi_{\{k^\ait\}}(\eta;\bm{\lambda}-k\tilde{\bm{\delta}}^\ai)
  \ \ (k\ge1),
  \label{1equiv}\\
  &\Xi_{\{m^\ai,\ldots,(k+m)^\ai\}}
  (\eta;\bm{\lambda}-m\tilde{\bm{\delta}}^\ait)\n
  &\sim\Xi_{\{(k+1)^\ait,\ldots,(k+m)^\ait\}}
  (\eta;\bm{\lambda}-(k+1)\tilde{\bm{\delta}}^\ai)
  \quad(k\ge1).
  \label{2equiv}
\end{align}
The simplest, $k=1$, of \eqref{1equiv} corresponds to the well known
fact that the $X_1$ type I and II polynomials are identical for L and J
\cite{os19}.
Deleting the first $k$ consecutive excited states, like the l.h.s.\ of
\eqref{1equiv} has led to many interesting phenomena in the ordinary
and discrete quantum mechanics \cite{gos,os15,os22}.
Classification of the equivalent classes leading to the same polynomials
is a challenging future problem.

The exceptional $X_\ell$ orthogonal polynomials of type I and II,
\cite{gomez,quesne,os16,os19,hos,stz} correspond to the simplest cases
of one virtual state deletion of that type, $\mathcal{D}=\{\ell^\ai\}$
or $\{\ell^\ait\}$, $\ell\ge1$:
\begin{align}
  \xi_{\ell}(\eta;\bm{\lambda})
  &=\Xi_{\mathcal{D}}(\eta;\bm{\lambda}+\ell\bm{\delta}+\tilde{\bm{\delta}}),
  \\
  P_{\ell,n}(\eta;\bm{\lambda})
  &=P_{\mathcal{D},n}(\eta;\bm{\lambda}+\ell\bm{\delta}+\tilde{\bm{\delta}})
  \times A,
\end{align}
where $\tilde{\bm{\delta}}=\tilde{\bm{\delta}}^{\ai,\ait}$ and
the multiplicative factor $A$ is $A=-1$ for XL1, $(n+g+\frac12)^{-1}$ for XL2,
$2(n+h+\frac12)^{-1}$ for XJ1 and $-2(n+g+\frac12)^{-1}$ for XJ2.
Most formulas between \eqref{hamD} and \eqref{fuchs} look almost the same
as those appearing in the theory of the exceptional orthogonal polynomials
\cite{os16,os17,os19,hos,stz,os20,os21,os23}.

In a recent paper \cite{gomez3} G\'{o}mez-Ullate et al discussed
``two-step Darboux transformations,'' which corresponds to the examples of
$\mathcal{D}=\{m_1^\ait, m_2^\ait\}$ for L in our scheme.
This paper stimulated the present work.
 
%%%%%%%%%%%%%%%%%%%%%%%%%%%%%%%%%%%%%%%%%%%%%%%%%%%%%%%%%%%%%%%
%                                                             %
%  4. Summary and Comments                                    %
%                                                             %
%%%%%%%%%%%%%%%%%%%%%%%%%%%%%%%%%%%%%%%%%%%%%%%%%%%%%%%%%%%%%%%
\section{Summary and Comments}
\label{summary}

This is a first short report on the discovery of infinite families of
multi-indexed orthogonal polynomials. They are obtained as the main part
of exactly solvable quantum mechanical systems, which are deformations
of the radial oscillator and the Darboux-P\"oschl-Teller potentials.
Although these new polynomials start at degree $\ell$ ($\ge1$), they
form a complete set of orthogonal polynomials.
The exactly solvable modified Hamiltonian and the eigenfunctions are
obtained by applying the modified Crum's theorem due to Adler to the
original system to delete a finitely many virtual state solutions of
type I and II. The simplest case of one virtual state deletion reproduces
the exceptional orthogonal polynomials \cite{gomez,quesne,os16,os19,hos}.
For the Jacobi case, the new polynomials provide infinitely many examples
of global solutions of second order Fuchsian differential equations with
$3+\ell$ regular singularities \cite{hos}. Like the undeformed theory,
the Laguerre results can be obtained from the Jacobi results in certain
confluence limits. But we presented all results in parallel, for better
understanding of the structure. The parameter ranges \eqref{Lbound} and
\eqref{Jbound} are conservative sufficient conditions with which the
proper boundary behaviours and the completeness are guaranteed.

As a first report, only the basic results are presented.
Some important issues cannot be included due to space restrictions,
for example, an alternative proof of shape invariance, the action of
various intertwining operators, etc.
Many important problems are remaining to be clarified.
For example, the bispectrality \cite{stz}, generating functions \cite{hos},
properties of the zeros of the denominator polynomial $\Xi_{\mathcal D}$
and the extra zeros of the multi-indexed polynomials $\{P_{{\mathcal D},n}\}$
\cite{hs4}, classification of the equivalent classes leading to the same
new polynomials.

The exceptional Wilson, Askey-Wilson, Racah and $q$-Racah polynomials
were constructed in the framework of discrete quantum mechanics 
\cite{os17,os20,os23,os24}, and the Crum-Adler theorem for discrete
quantum mechanics were also presented \cite{os15,gos,os22}.
The method of virtual states deletion presented in this Letter is
applicable to discrete quantum mechanics and it is easy to write down
general formulas like as \eqref{mnham}--\eqref{finpot}.
This gives multi-indexed orthogonal polynomials corresponding to the
deformations of the Wilson, Askey-Wilson, Racah and $q$-Racah polynomials.
In concrete examples, more than two types of virtual state solutions
are available and much richer structures are expected.
We will report on these topics elsewhere.

Before closing this Letter, let us make two small comments.
The original Schr\"{o}dinger equation \eqref{sheq} has a general solution
for an arbitrary energy $\mathcal{E}$, $\mathcal{H}\psi(x)=\mathcal{E}\psi(x)$. 
If we define the corresponding $[M,N]$ deletion solution
\begin{equation*}
  \psi^{[M,N]}(x)\eqdef
  \frac{\text{W}[\tilde{\phi}_{d_1}^\ai,\ldots,\tilde{\phi}_{d_M}^\ai,
  \tilde{\phi}_{d_1}^\ait,\ldots,\tilde{\phi}_{d_N}^\ait,\psi](x)}
  {\text{W}[\tilde{\phi}_{d_1}^\ai,\ldots,\tilde{\phi}_{d_M}^\ai,
  \tilde{\phi}_{d_1}^\ait,\ldots,\tilde{\phi}_{d_N}^\ait](x)},
\end{equation*}
it solves the deformed Schr\"{o}dinger equation with the same energy, too
\begin{equation*}
  \mathcal{H}^{[M,N]}\psi^{[M,N]}(x)=\mathcal{E}\psi^{[M,N]}(x).
\end{equation*}
Among the multi-indices, we can include a finite number of continuous
real parameters as labeling the type I virtual solutions as described in
\cite{os21,junkroy,duttaroy}. The resulting orthogonal functions are no
longer polynomials, but most properties of the multi-indexed polynomials
are shared by these new orthogonal functions.

%%%%%%%%%%%%%%%%%%%%%%%%%%%%%%%%%%%%%%%%%%%%%%%%%%%%%%%%%%%%%%%
%                                                             %
%  Acknowledgments                                            %
%                                                             %
%%%%%%%%%%%%%%%%%%%%%%%%%%%%%%%%%%%%%%%%%%%%%%%%%%%%%%%%%%%%%%%
\section*{Acknowledgements}
We thank T.\,Oshima and Y.\,Haraoka for useful discussion on Fuchsian
differential equations.
R.\,S. is supported in part by Grant-in-Aid for Scientific Research
from the Ministry of Education, Culture, Sports, Science and Technology
(MEXT), No.22540186.

%%%%%%%%%%%%%%%%%%%%%%%%%%%%%%%%%%%%%%%%%%%%%%%%%%%%%%%%%%%%%%%
%                                                             %
%  References                                                 %
%                                                             %
%%%%%%%%%%%%%%%%%%%%%%%%%%%%%%%%%%%%%%%%%%%%%%%%%%%%%%%%%%%%%%%

\end{document}